\documentclass[twocolumn]{aastex7}
\usepackage{enumerate}
\usepackage{amssymb, amsmath}
\usepackage{natbib}
\usepackage{color}
\usepackage{ulem}
\usepackage{appendix}
\usepackage{url}
\usepackage{hyperref}
\usepackage[stable]{footmisc}
\usepackage{relsize}
\usepackage{mathtools}

\definecolor{red}{rgb}{1.0,0.0,0.0}

\begin{document}

\title{SCExAO/CHARIS High-Contrast Pre-Launch Vetting of \\Roman Coronagraph Technology Demonstration PSF Reference Stars\footnote{Based in part on data collected at Subaru Telescope, which is operated by the National Astronomical Observatory of Japan.}}

\email{thayne.currie@utsa.edu}
\author[0000-0002-7405-3119]{Thayne Currie}
\affiliation{Department of Physics and Astronomy, University of Texas at San Antonio, San Antonio, TX 78249, USA}
 \affiliation{Subaru Telescope, National Astronomical Observatory of Japan, 
650 North A`oh$\bar{o}$k$\bar{u}$ Place, Hilo, HI  96720, USA}
\email[show]{thayne.currie@utsa.edu}
\author[0000-0003-2426-3852]{Jie Li}
\affiliation{Department of Physics and Astronomy, University of Texas at San Antonio, San Antonio, TX 78249, USA}
\email[show]{jie.li@utsa.edu}
\author{Mona El Morsy}
\affiliation{Department of Physics and Astronomy, University of Texas at San Antonio, San Antonio, TX 78249, USA}
\email[]{mona.elmorsy@utsa.edu}
\author[]{Olivier Guyon}
 \affiliation{Subaru Telescope, National Astronomical Observatory of Japan,
650 North A`oh$\bar{o}$k$\bar{u}$ Place, Hilo, HI  96720, USA}
\affiliation{Astrobiology Center, 2-21-1 Osawa, Mitaka, Tokyo 181-8588, Japan}
\email[]{guyon@naoj.org}
\author{Julien Lozi}
 \affiliation{Subaru Telescope, National Astronomical Observatory of Japan,
650 North A`oh$\bar{o}$k$\bar{u}$ Place, Hilo, HI  96720, USA}
\email[]{lozi@naoj.org}
\author{Erica Dykes}
\affiliation{Department of Physics and Astronomy, University of Texas at San Antonio, San Antonio, TX 78249, USA}
\email[]{erica.dykes@my.utsa.edu}
\author{Danielle Bovie}
\affiliation{Department of Physics and Astronomy, University of Texas at San Antonio, San Antonio, TX 78249, USA}
\email[]{danielle.bovie@my.utsa.edu}
\author{Sebastien Vievard}
 \affiliation{Subaru Telescope, National Astronomical Observatory of Japan,
650 North A`oh$\bar{o}$k$\bar{u}$ Place, Hilo, HI  96720, USA}
\affiliation{Space Science and Engineering Initiative, College of Engineering, University of Hawai`i, Hilo, HI 96720, USA}
\affiliation{Institute for Astronomy, University of Hawaii, Hilo, HI 96720, USA}
\email[]{vievard@naoj.org}
\author{Garima Singh}
 \affiliation{Subaru Telescope, National Astronomical Observatory of Japan,
650 North A`oh$\bar{o}$k$\bar{u}$ Place, Hilo, HI  96720, USA}
\email[]{singh@naoj.org}
\author{Kyohoon Ahn}
\affiliation{Korea Astronomy and Space Science Institute, 776 Daedeok-daero, Yuseong-gu, Daejeon 34055, Republic of Korea}
 \affiliation{Subaru Telescope, National Astronomical Observatory of Japan,
650 North A`oh$\bar{o}$k$\bar{u}$ Place, Hilo, HI  96720, USA}
\email[]{ahn@naoj.org}
\author{Vincent Deo}
 \affiliation{Optical Sharpeners, Manosque, France}
 \affiliation{Subaru Telescope, National Astronomical Observatory of Japan,
650 North A`oh$\bar{o}$k$\bar{u}$ Place, Hilo, HI  96720, USA}
\email[]{vdeo@naoj.org}
\author{Yoshito Ono}
 \affiliation{Subaru Telescope, National Astronomical Observatory of Japan,
650 North A`oh$\bar{o}$k$\bar{u}$ Place, Hilo, HI  96720, USA}
\email[]{yono@naoj.org}

\shortauthors{Currie, Li, El Morsy et al. 2026}
\begin{abstract}
We present deep, SCExAO/CHARIS high-contrast integral field spectroscopy and archival imaging of four candidate Roman Coronagraph PSF reference stars within/near the Roman Continuous Viewing Zone and potentially suitable for the Coronagraph's key technology demonstration targets HIP 71618 and HIP 54515.  For CHARIS data, we achieve 5-$\sigma$ contrasts down to $\sim$1.4$\times$10$^{-5}$, $\sim$6$\times$10$^{-6}$,  and 10$^{-6}$ to 4$\times$10$^{-7}$ at 0\farcs{}16, 0\farcs{}25, and 0\farcs{}5 to 1\arcsec{}.  Companion mass limits rule out brown dwarfs at $\rho$ $\sim$ 0\farcs{}15--0\farcs{}25 and massive planets at wider separations around all targets.  More critically, for three of the four references our analysis disfavors companions with $V$ band contrasts brighter than 10$^{-8}$, 10$^{-9}$, and $10^{-10}$ at 0\farcs{}15, 0\farcs{}3, and 1$\arcsec{}$.  Unless these targets have faint substellar companions within $\rho$ $\sim$ 0\farcs{}15, they likely lack background stars or companions that could corrupt the Roman Coronagraph's dark hole digging to preclude detecting reflected-light planets.  For $\alpha$ Cep,  our limits are a factor of $\sim$10 worse but still meet the TTR5 limit of 10$^{-7}$ beyond $\rho$ $\sim$ 0\farcs{}25: beyond 0\farcs{}4, they exclude a Jupiter-twin reflected-light companion (10$^{-9}$).  Archival Keck/NIRC2 data likewise find no substellar companions with $\Delta$V $>$ 10$^{-8}$ at wider separations.   Finally, we assess the observability of HIP 71618 and HIP 54515  -- updated for Roman's launch date of August 30, 2026.   Adding $\gamma$ Boo -- not currently in the Roman CPP team reference-star list -- would improve schedulability for the tech demo's key targets.

\end{abstract}

\section{Introduction}
The Roman Space Telescope Coronagraphic Instrument (CGI) technological demonstration (hereafter, the ``tech demo") is a critical stepping stone towards imaging and characterizing an Earth-like planet around a nearby Sun-like star with the Habitable Worlds Observatory \citep{astro2020,Currie2023b}.
NASA’s core criterion for the CGI tech demo's success is a Threshold Technical Requirement (TTR5) that CGI must be able to detect at $>$ 5-$\sigma$ a point source at a contrast of $\le$10$^{-7}$ at  $\lambda$ $<$ 600 nm ($>$ 10\% bandpass) located 6-9 $\lambda$/D ($\sim$0\farcs{}3-0\farcs{}45) from a V $\le$ 5 star \footnote{See \url{https://roman.gsfc.nasa.gov/science/rsig/2021/Roman_Requirements_20201105.pdf} and \url{https://roman.ipac.caltech.edu/docs/CGI_info_talks/day_oct26/Hildebrandt.pdf}}.  
Tech demo observations will occur during the first 18 months of Roman's full operation.  They
involve slewing to a bright (V $<$ 3) reference star, modulating the CGI deformable mirrors (DMs) to dig a dark hole (DH), slewing to a nearby $V$ $\lesssim$ 5 target star\footnote{CGI achieving its threshold performance on a star fainter than V = 5, while more challenging, meets TTR5 (JPL CGI team, pvt. comm, 2020).  CGI should achieve a DH of good depth at V = 6 and of some depth down to V $\sim$ 7 \citep{Shi2017}.}, and applying the DM shape from the reference to this target.  This sequence is repeated multiple times.

Beyond achieving TTR5, CGI has five technology Objectives, two of which\footnote{\url{https://roman.gsfc.nasa.gov/science/rsig/2021/Roman\_Requirements\_20201105.pdf}} require detecting companions ``at a contrast level and separation [requiring] a functional coronagraph and wavefront control capability" (2.2.1; "Coronagraph with Active Wavefront Control") and characterizing ``photometry, spectroscopy, and astrometry" of at least one of them (2.2.5; "High-Contrast Data Processing").  A third focuses on performance characterization of CGI (2.2.4; "High Contrast Performance Characterization ").

To fulfill the core TTR5 requirement and these Objectives,
The Roman Coronagraph Community Participation (CPP) Program \citep{Wolff2024} has now recommended a suite of tech demo target observations\footnote{\url{http://conference.ipac.caltech.edu/SpiritofLyot6/slides/Wednesday\_Morning\_Wolff\%20-\%20Schuyler\%20Wolff.pdf}}.  Among these observations, CGI 575nm targeting of HIP 71618 -- a bright A star hosting a substellar companion at $\rho$ $\sim$ 0\farcs{}3 \citep{ElMorsy2025} will demonstrate TTR5 in the post-commissioning phase.   Spectroscopic mode observations of HIP 71618 and HIP 54515, an A star orbited by a superjovian planet at $\rho$ $\sim$ 0\farcs{}2 \citep{Currie2026a}, will fulfill multiple Objectives\footnote{The HIP 71618 B and HIP 54515 b companions were both discovered from the Observing Accelerators with SCExAO Imaging Survey (OASIS), supported by NASA Headquarters to identify suitable Roman Coronagraph technology demonstration targets (PI T. Currie, Co-I M. Kuzuhara) \citep{ElMorsy2024a}.  In particular, the HIP 71618 system discovered by M. El Morsy (\citeyear{ElMorsy2025}) provides the critical TTR5-achieving target for the Roman Coronagraph. }\footnote{The CPP team slide text typographical error notwithstanding, the current peer-reviewed literature  decisively interprets HIP 54515 b as a planet, not a brown dwarf \citep{Currie2026a}.}.  Multiple candidate PSF reference stars are schedulable in combination with observations of HIP 71618 and HIP 54515, including compelling ones that were not included in the Roman CPP team's initial candidate PSF reference identification \citep{Currie2025a,Currie2025b,Currie2026b}.  

Properly vetting candidate PSF reference stars is a critical preparatory activity for achieving TTR5 and fulfilling these Objectives with HIP 71618, HIP 54515, and other targets.  If the reference star is too faint, then CGI would require an excessive amount of time to dig a dark hole. If the star is partially resolved then CGI's  performance is compromised; a faint companion within CGI's control region or a bright one at wide separations also degrade the instrument's ability to dig a dark hole.  Quantitatively, a reference star a) should have V $\lesssim$ 3\footnote{This brightness threshold is driven by the time needed to dig a dark hole.   The V $\lesssim$ 3 requirement is different than the brightness requirement for a \textit{target} star whose DH is generated from reference star observations.}, b) should have an angular size $\lesssim$2 mas, and c) must lack companions capable of compromising CGI's DH digging at TTR5-relevant levels ($\gtrsim$ 10$^{-7}$--10$^{-8}$ contrast) \citep{Hom2026}.

The Roman spacecraft's thermal environment and pointing on the sky also limits potential reference stars for a given tech demo target and drives schedulability.  Roman will orbit at the Sun-Earth L2 point.  Ensuring a common spacecraft thermal environment between reference star and tech demo targets maximizes CGI's achievable contrast on the latter targets.  Simulations suggest that performance will degrade at pitch angle differences greater than $\sim$5$^{o}$ \citep{Hom2026}.  At any given time, $\sim$41\% of the sky is within a ``keep out zone" where light primarily from the Sun would preclude Roman observations.   A primary scientific focus of Roman during the tech demo window -- long duration, high-cadence extragalactic science (e.g. the High-Latitude Time-Domain Survey) -- requires continuous sky access, which is only possible within Roman Continuous Viewing Zone (CVZ) region more than 54$^{o}$ from the ecliptic plane \citep{Rose2023}.   
PSF reference stars and tech demo targets within or near the CVZ thus maximize the schedulability of CGI observations, as they can be more flexibly slotted in during gaps in these cadences.  

Recently, \citet{Hom2026} provided the first systematic vetting of the full CPP team reference star list, analyzing optical interferometry, speckle interferometry, and shallow adaptive optics (AO) to identify contaminating companions.  Their data achieved typical contrasts of 6-7 mags in the near-IR at $\rho$ $\sim$ 0\farcs{}6.  While these data confirmed that the candidate PSF references to HIP 71618 and HIP 54515 lack bright binaries, these limits translate into contrasts of $\gtrsim$ 10$^{-4}$ in $V$ band: a factor of 1000 brighter than the limit needed to vet the references for TTR5 suitability.   To constrain the existence of companions at a near-IR depth that translates into $<$ 10$^{-7}$ at optical wavelengths, extreme AO imaging capable of detecting companions at 10$^{-5}$--10$^{-6}$ at 0\farcs{}1--1\arcsec{} are needed.

In this paper, we present deep SCExAO/CHARIS imaging and archival imaging of candidate PSF reference stars suitable for pairing with CGI's highest priority tech demo targets -- HIP 71618 and HIP 54515 -- and located within the Roman CVZ.  Section \ref{sec:data} describes our observations, data reduction, and PSF-subtracted images.  We then derive contrast curves and convert these limits into mass limits in Section \ref{sec:results}.  Finally, given Roman's updated launch schedule (August 30, 2026), we assess the observability of HIP 71618 and HIP 54515 as a part of the technology demonstration phase (Section \ref{sec:schedule}).

\begin{deluxetable*}{lllllllll}[h]
     \tablewidth{0pt}
    \tablecaption{Target List}
    \tablehead{\colhead{Target} & \colhead{Alt. Name} &  \colhead{V} & \colhead{Spectral Type} & \colhead{Distance (pc)} & \colhead{Age (Myr)} & \colhead{Roman CPP Team Rank} & \colhead{Current Limits$^{a}$} & \colhead{References$^{b}$}}
    \startdata
   $\eta$ UMa& HD 120315& 1.86& B3V & 31.87 & 10 & A/B & early M & 1,2,3\\
    $\alpha$ Cep & HIP 105199  & 2.46& A8Vn & 15.04 & 900 & B & mid M & 4, 5, 2, 6\\
    $\epsilon$ UMa  & HD 112185 & 1.77& A1III-IVpkB9 & 25.31 & 530 & B/C & early M & 4, 2, 7\\
    $\gamma$ Boo& HD 127762  & 3.02& A7IV+(n)  & 26.40 & 900 & -- & --&  8, 2, 9\\
    \enddata
    \tablecomments{a) Contrast limits are defined at an angular separation of 0\farcs{}6.  b) References are: 1) \citet{MorganKeenan1973}, 2) \citet{vanLeeuwen2007}, 3) \citet{Tetzlaff2011}, 4) \citet{Gray2003}, 5) \citet{Oja1991}, 6) \citet{DavidHillenbrand2015}, 7) \citet{BrandtHuang2015}, 8) \citet{Ventura2007}, and 9) \citet{Gray2001}.  
   }
    \label{targetlist}
    \end{deluxetable*}

\begin{deluxetable*}{lllllllll}[ht]
     \tablewidth{0pt}
    \tablecaption{Observing Log\label{obslog}}
    \tablehead{\colhead{UT Date} & \colhead{Target} & \colhead{Alt. Name} &  \colhead{Instrument} & \colhead{Passband} & \colhead{Seeing (\arcsec{})} 
    & \colhead{$t_{\rm exp}$ (s)} & \colhead{$N_{\rm exp}$} & \colhead{$\Delta$PA ($^{o}$)} }
    \startdata
    \textbf{New Data}\\
    20260427 & $\eta$ UMa& HD 120315$^{a}$& AO3k+SCExAO/CHARIS&$JHK$ &0.4-0.6 & 30.98 & 166 & 35.85  \\
    20260427 & $\alpha$ Cep & HIP 105199$^{b}$  & AO3k+SCExAO/CHARIS&$JHK$ &0.6-0.8 & 10.13 & 126 & 10.13  \\
    20260524 & $\epsilon$ UMa  & HD 112185 & AO3k+SCExAO/CHARIS&$JHK$ &0.3-0.4 & 20.65 & 240 & 39.22 \\
    20260524 & $\gamma$ Boo& HD 127762  & AO3k+SCExAO/CHARIS&$JHK$ &0.3-0.4 & 41.31 & 140 & 55.50 \\
    \textbf{Archival Data}\\
    20120904 & $\alpha$ Cep & HIP 105199 & Keck II/NIRC2& $K_{\rm s}$ & 0.8-1.2 & 30 & 60 & 15.19 \\
    20140410 & $\eta$ UMa& HD 120315 & Keck II/NIRC2& $K_{\rm s}$ & 0.5-0.6 & 1.08, 2.19, 10 & 34 & 5.74 \\
    \enddata
    \tablecomments{$JHK$ refers to the CHARIS broadband mode which covers 1.16--2.37 $\mu m$.  a) data suffered moderate or slight ``low-wind effect" \citep{Milli2018} degrading raw contrast by a factor of 2--3 over half the field of view.  
    b) Data suffered significant ``low-wind effect" degrading raw contrast by more than a factor of 10 over half the field of view.
   }
    \label{obslog_scexao}
    \end{deluxetable*}
    
\section{Data}
\label{sec:data}
\subsection{Observations}

Table \ref{obslog_scexao} lists our targets and details about their observations, obtained as part of an approved Subaru-Gemini Time Exchange program focused on Roman Coronagraph target vetting (S26A-TE208-RG; PI T. Currie).  To select targets, we focused on candidate PSF reference stars identified from the Roman CPP located in or near the Roman CVZ and potentially suitable (ranked as ``A", ``B", or ``C" as of 2026 June 18) for Band 1 (575 nm) observations with the Hybrid Lyot coronagraph\footnote{\url{https://docs.google.com/spreadsheets/d/1p5r0VmjBCjXU25daJl5oJOPoPh1V79nuESbnwmca0s0/edit?gid=1677191164\#gid=1677191164}}.  We further downselected to stars visible in the northern sky that could conceivably be used as references for HIP 71618 and HIP 54515.  Additionally, we consider $\gamma$ Boo, which is not in the CPP list but was identified as a potential candidate reference star in \citet{Currie2026b} given the Roman CPP criteria\footnote{While $\gamma$~Boo ($V = 3.02$) marginally exceeds the nominal $V \lesssim 3$
guideline, the brightness criterion for reference star selection is set by dark-hole digging time.  This star is $\sim$ 1.7 times fainter than $\alpha$ Cep and just $\sim$ 3\% fainter than faintest candidate reference stars considered by the Roman CPP team.   If the digging time scales with photon noise, this
lengthens the required integration by $\sim$2.8$\times$ relative to $\alpha$ Cep and 6\% compared to the faintest candidate reference star ---a modest scheduling
cost.     A V=3.02 star is then  well within the depth at which
CGI still digs a usable dark hole within a feasible time.  Our analysis shows that other properties of $\gamma$ Boo determined in this study -- e.g. $V$-band contrast limits and schedulability -- make it a well-suited PSF reference star. }

The final sample consists of four stars: $\eta$ UMa, $\alpha$ Cep, $\epsilon$ UMa, and $\gamma$ Boo (Table \ref{targetlist}).  The first three of these stars were already vetted for bright binaries in \citet{Hom2026}.  $\gamma$ Boo is listed as a tight binary from 1970s-era speckle interferometry \citep{Morgan1978}.   However, the star is a $\delta$ Scu variable whose pulsations could be mistaken for autocorrelation peaks in speckle data.  Prior non-coronagraphic high-contrast imaging data from the LEECH survey fail to resolve a bright binary companion around this star \citep{Stone2018}; the satellite spots for our CHARIS data presented in this work below likewise fail to reveal evidence for a bright companion.  Thus, the prior claim of a binary in this system is plausibly spurious.

Using AO3k~+SCExAO~/CHARIS in its low spectral resolution (broadband) mode  covering the $JHK$ passbands simultaneously (1.16-2.37 $\mu$m at $\mathcal{R}$ $\sim$ 18), we observed the stars on 2026 04 27 and 2026 05 24.  The $\gamma$ Boo data is had the highest quality and deepest raw contrast.  For the April 2026 data, the AO residuals left low-order aberrations consistent with low-wind effect \citep{Milli2018}, degrading the raw contrast by a factor of $\sim$2--3 for $\eta$ UMa and $\gtrsim$10 for $\alpha$ Cep relative to $\gamma$ Boo.  The $\eta$ UMa data had a visible wind-driven halo likewise degrading the raw contrast.
The May 24 data for $\gamma$ Boo lacked all of these residuals, leaving a 360-degree deep dark hole.  For all data sets, we used a beamsplitter to send only 10\% of the near-infrared (IR) light to CHARIS to prevent saturation or nonlinearity.

We obtained all data in pupil tracking mode to enable angular differential imaging \citep[ADI;][]{Marois2006}, yielding parallactic angle rotations of $\sim$ 10--55 degrees (1--5.5 $\lambda$/D at 1.65 $\mu m $ and $\rho$ $\sim$ 0\farcs{}25).  Total exposure times ranged between 21 minutes and 96 minutes.  For all data, we used a Lyot coronagraph with a 0\farcs{}113-radius occulting spot.     For astrometric and spectrophotometric calibration, we modulated the SCExAO deformable mirror with a 25 nm amplitude to generate satellite spots \citep[][]{Jovanovic2015-astrogrids}.  

To these data, we added archival Keck/NIRC2 $K_{\rm s}$ band imaging of $\alpha$ Cep and $\eta$ UMa from open use science programs (PIs B. Macintosh, E. Nielsen) 
 downloaded from the Keck Observatory Archive. 
Visual inspection of the NIRC2 data showed that they were typically saturated out to $\rho$ $\sim$ 0\farcs{}75. 
These data are thus primarily useful for setting limits on companions beyond the CHARIS outer working angle ($\rho$ $\approx$ 1\farcs{}1).


\subsection{Data Reduction}
For all CHARIS data, we used the pipeline from \citet{Brandt2017} to extract data cubes from the raw exposures as in previous work\footnote{\url{https://github.com/PrincetonUniversity/charis-dep}}.  
The CHARIS Data Processing Pipeline (CHARIS DPP) \citep{Currie2020a}\footnote{\url{https://github.com/thaynecurrie/charis-dpp}} performed subsequent data reduction steps -- sky subtraction, image registration, spectrophotometric calibration, and spatial filtering -- following prior work \citep[e.g.][]{Currie2023b, Currie2026a,Bovie2025}.\footnote{All previous CHARIS publications analyzed data taken before the October 2025 repositioning of SCExAO/CHARIS behind the Nasmyth beamswitcher (NBS).   Engineering tests showed that the NBS installation induced a parity flip and substantial north position angle offset on the CHARIS detector compared to its pre-NBS astrometric calibration.   We recalibrated the CHARIS astrometric solution using observations of binary companions in November 2025 (e.g. HD 1160) and modified the CHARIS DPP to account for both the parity flip and new north position angle offset.   While a future paper will report a precise value for the new north position angle offset, we find an estimated additional offset of $\sim$78.3$^{o}$ must be added to the parallactic angle to recover the prior astrometric solution, which found a north position angle offset of -2.03$^{o}$ \citep{Currie2026a}.  }
For CHARIS point-spread function (PSF) subtraction, we used the Adaptive, Locally Optimized Combination of Images algorithm in combination with ADI and then spectral differential imaging (SDI) on the post-ADI residuals \citep[A-LOCI][]{Currie2012,Currie2015}.   

To improve our achievable contrast, we explored the A-LOCI free parameter space, iteratively processing the data with different algorithm parameter space, computing attenuation maps of signal loss over the entire field, and then producing contrast curves\footnote{The parameter space exploration was made efficient by significant updates to the CHARIS DPP, removing latent file I/O in favor of pointer arrays, replacing the standard LAPACK-sourced truncated SVD calculation with a faster, equivalent operation using eigendecomposition -- given that the covariance matrix is a positive semi-definite (Gram) matrix -- and parallelizing the PSF subtraction and signal loss computation over available cores.  We confirmed equivalent results with the earlier-era CHARIS DPP.  The PSF subtraction steps are consequentially factors of $\sim$10-20 faster than before on a modern laptop with $>$10 cores, with an even larger speed gain for the full subtraction, attenuation map, and contrast curve procedure described above.  The upgraded CHARIS DPP will incorporate additional improvements - e.g. forward-model matched filtering \citep{Ruffio2017}.  It will be described in full in a later work.}.  We considered varying the rotation gap criteria ($\delta$) and optimization area sizes (i.e. the area over which a reference PSF is computed; $N_{A}$) 
\citep[see also][]{Lafreniere2007,Marois2010b,Currie2012,Currie2026a}. 
To correct for signal loss due to processing \citep[e.g.][]{Marois2010b,Pueyo2016}, we employed the forward-modeling approach described in \citet{Currie2018}.   The balance of signal loss vs. speckle suppression with SDI may vary depending on the assumed planet spectrum \citep[e.g. L dwarf vs. T dwarf;][]{Marois2014}.  Thus, for each target, we performed two forward-modeling steps: one assuming a cloudy low-gravity L0 dwarf spectrum and another adopting a T2 dwarf spectrum. Typically, we achieved our best contrasts over a parameter space focused on $\delta$ $\sim$ 0.3--0.6 and $N_{\rm A}$ $\sim$ 50-100.   

 For the NIRC2 data, we used the ADI-based pipeline from \citet{Currie2011}, following reduction steps as described in \citet{ElMorsy2024b,ElMorsy2025}, including image registration, photometric calibration, spatial filtering, and PSF subtraction.  We adopted the same forward-modeling framework as with the CHARIS data.

 \section{Results}\label{sec:results}

 \begin{figure*}[ht!]

\includegraphics[width=0.5\textwidth]{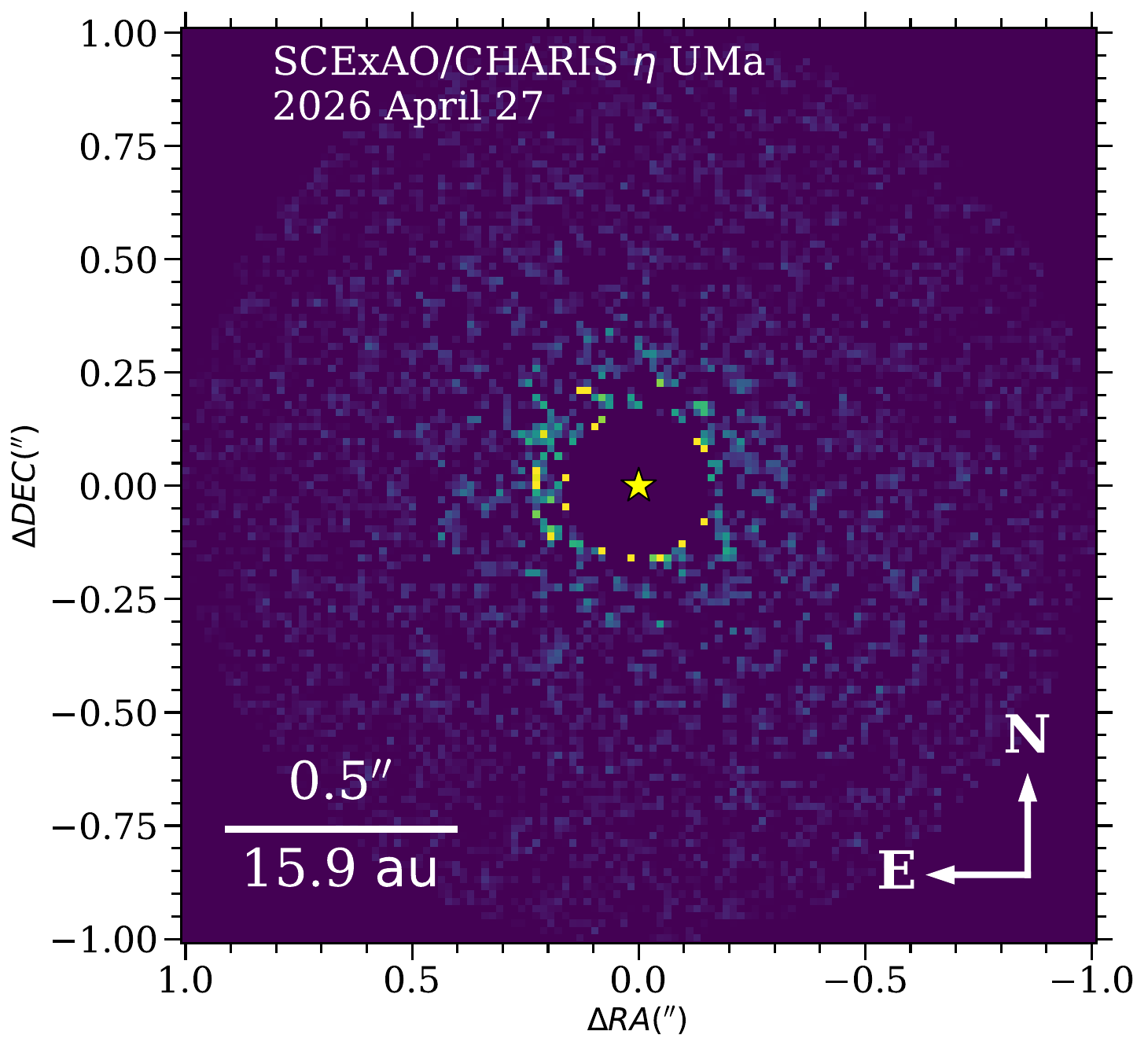}
\includegraphics[width=0.5\textwidth]{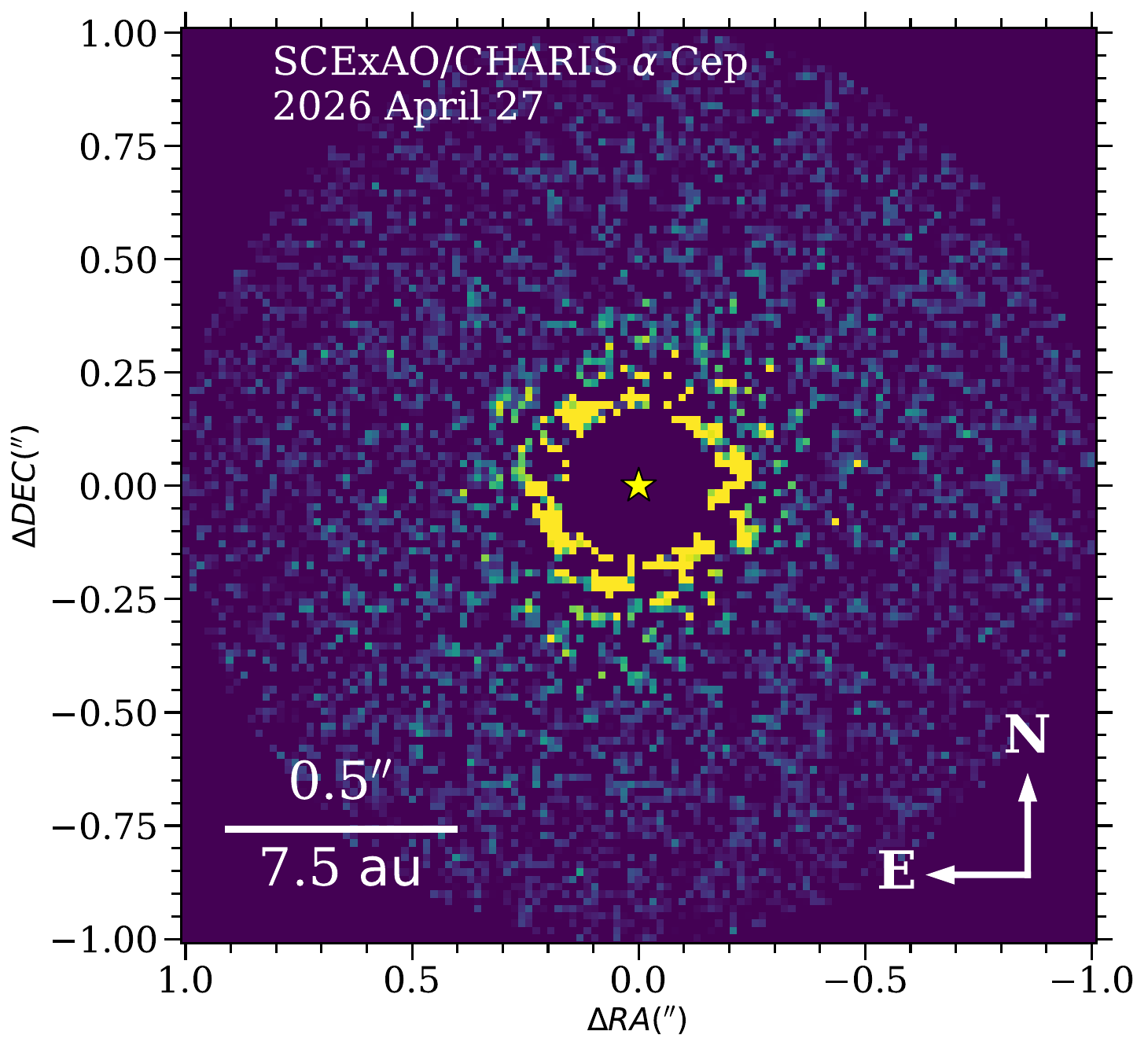}\\
\includegraphics[width=0.5\textwidth]{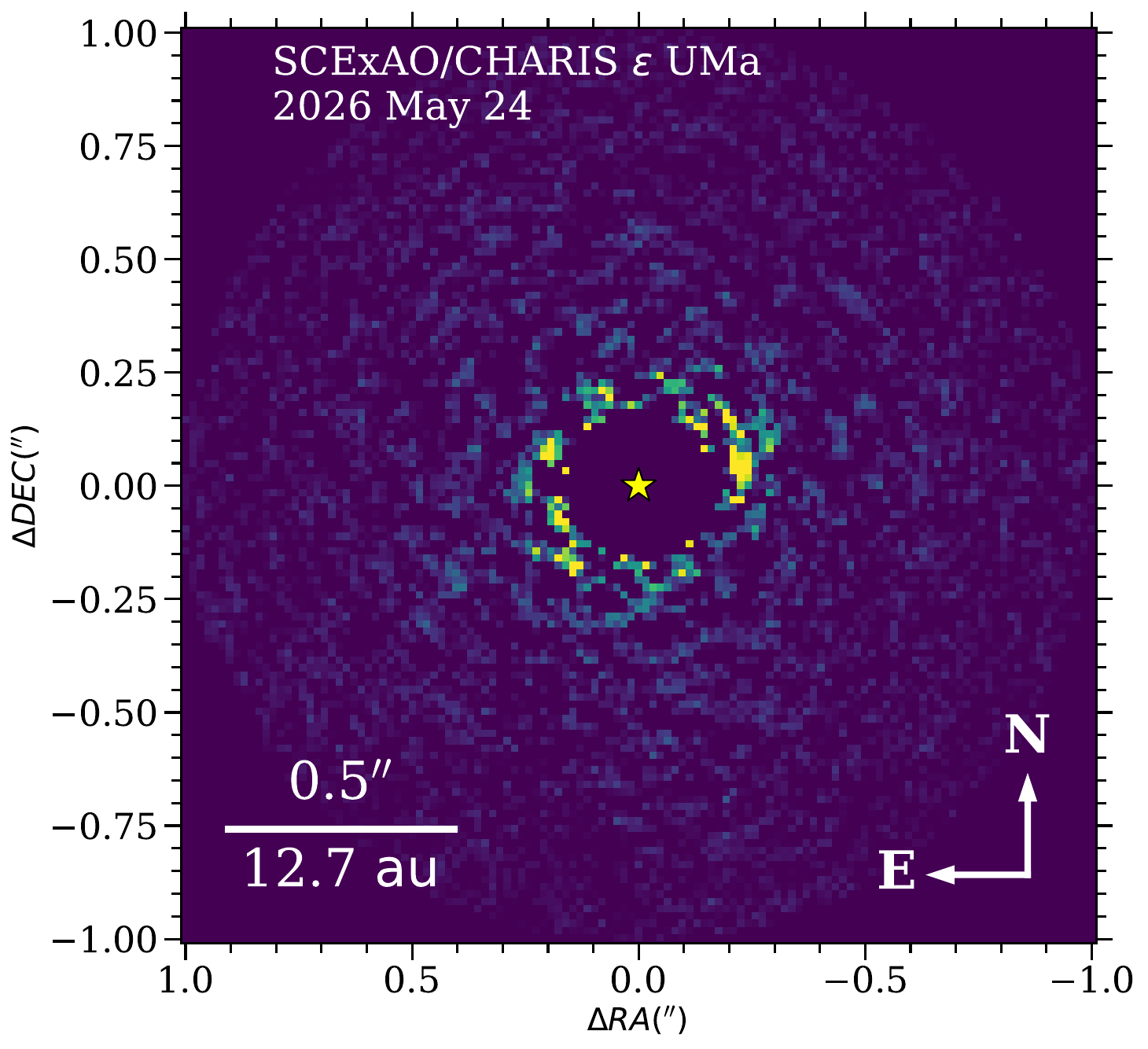}
\includegraphics[width=0.5\textwidth]{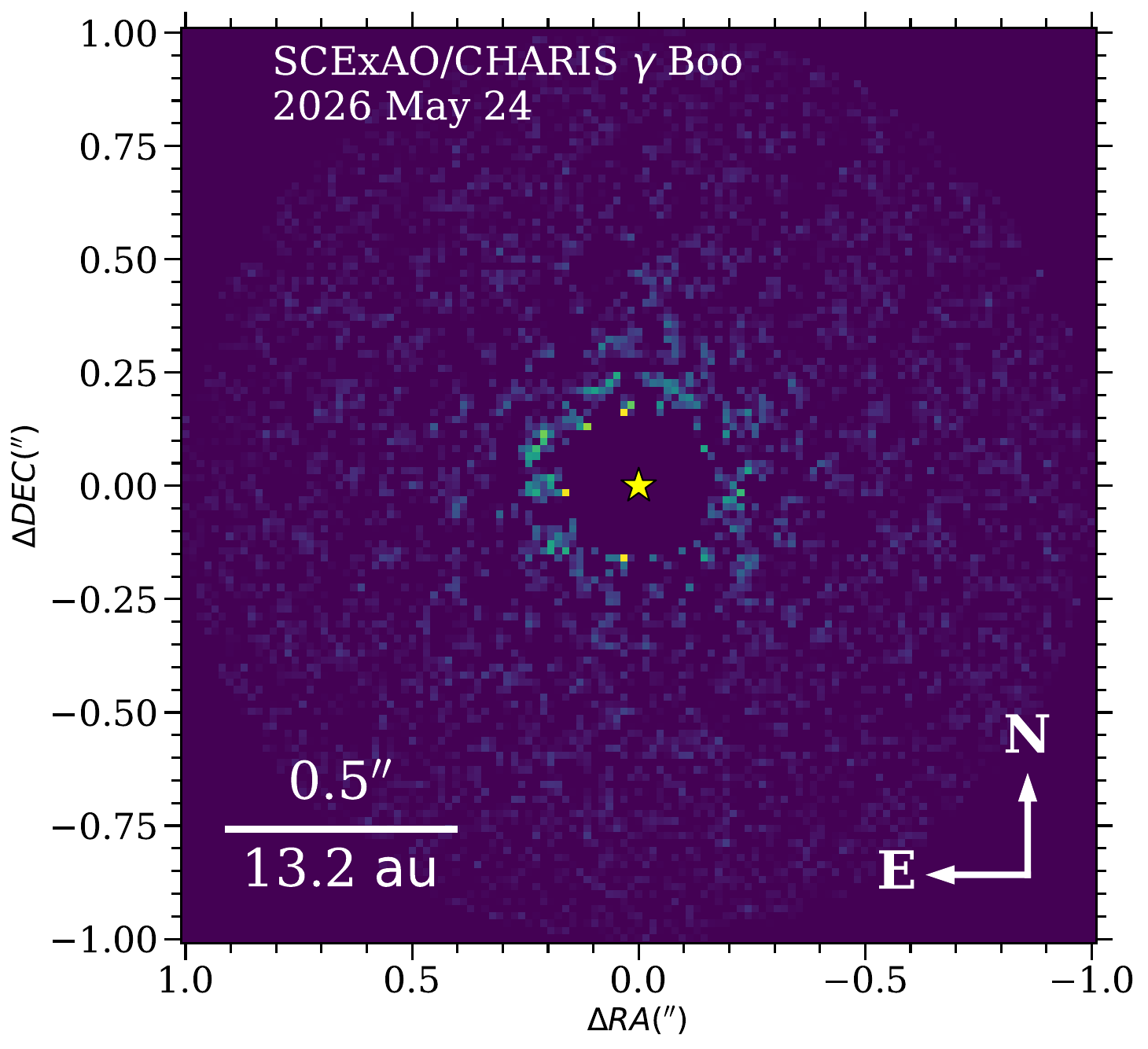}
   \vspace{-0.3in}
  \caption{ASDI-reduced CHARIS images for (clockwise from top left) $\eta$ UMa, $\alpha$ Cep,  $\gamma$ Boo, and $\epsilon$ UMa.  The color stretch is normalized to show the same contrast range -- $\approx$ 0--10$^{-6}$, without accounting for self-subtraction--for each target. Typical post-PSF subtraction throughput near the CHARIS coronagraph mask edge ($\rho$ $\approx$ 0\farcs{}15--0\farcs{}3) is 20-35\%.}
  \label{fig:finalimages}
\end{figure*}


 \begin{figure}[ht!]

\includegraphics[width=0.5\textwidth]{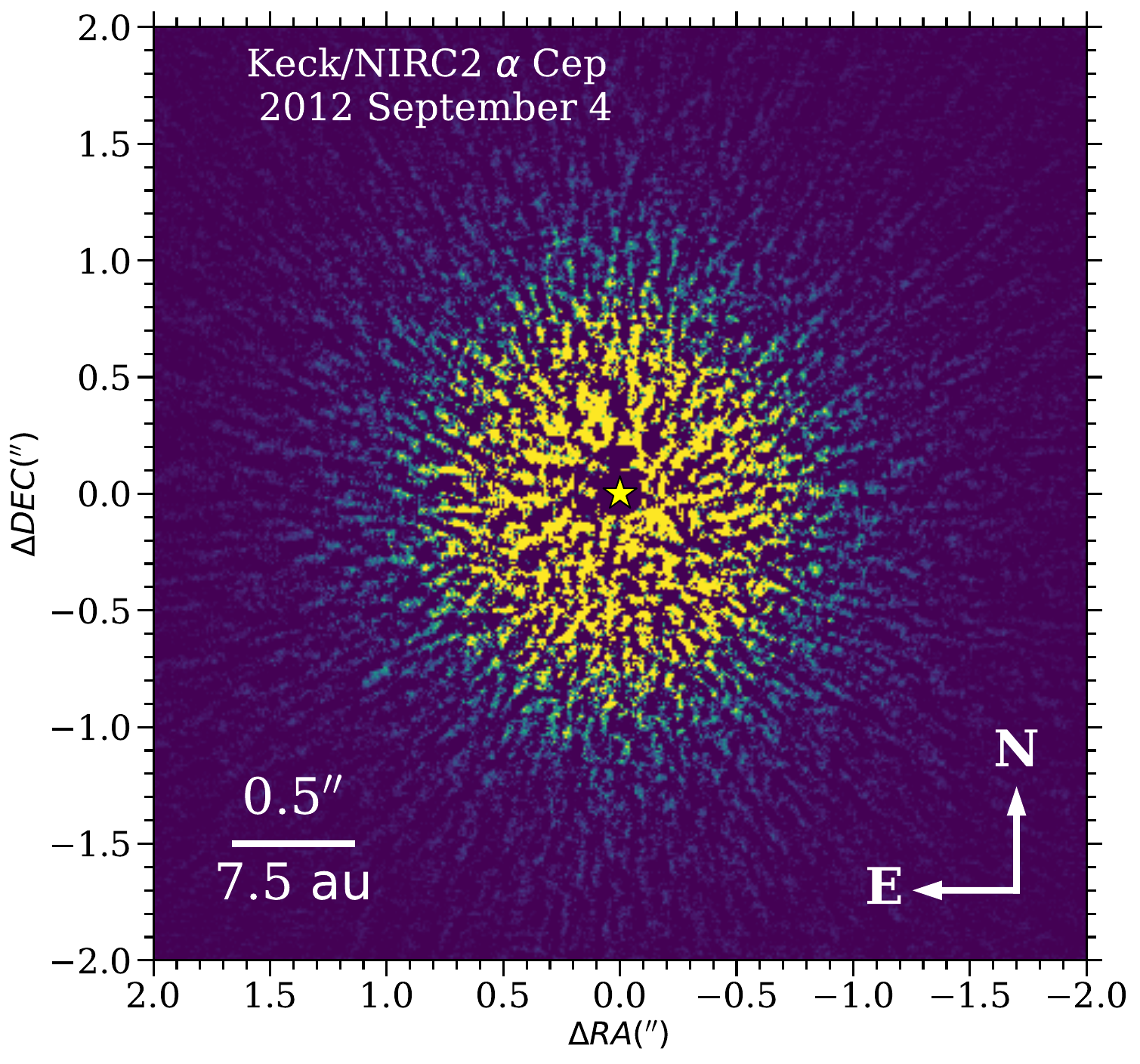}\\
\includegraphics[width=0.5\textwidth]{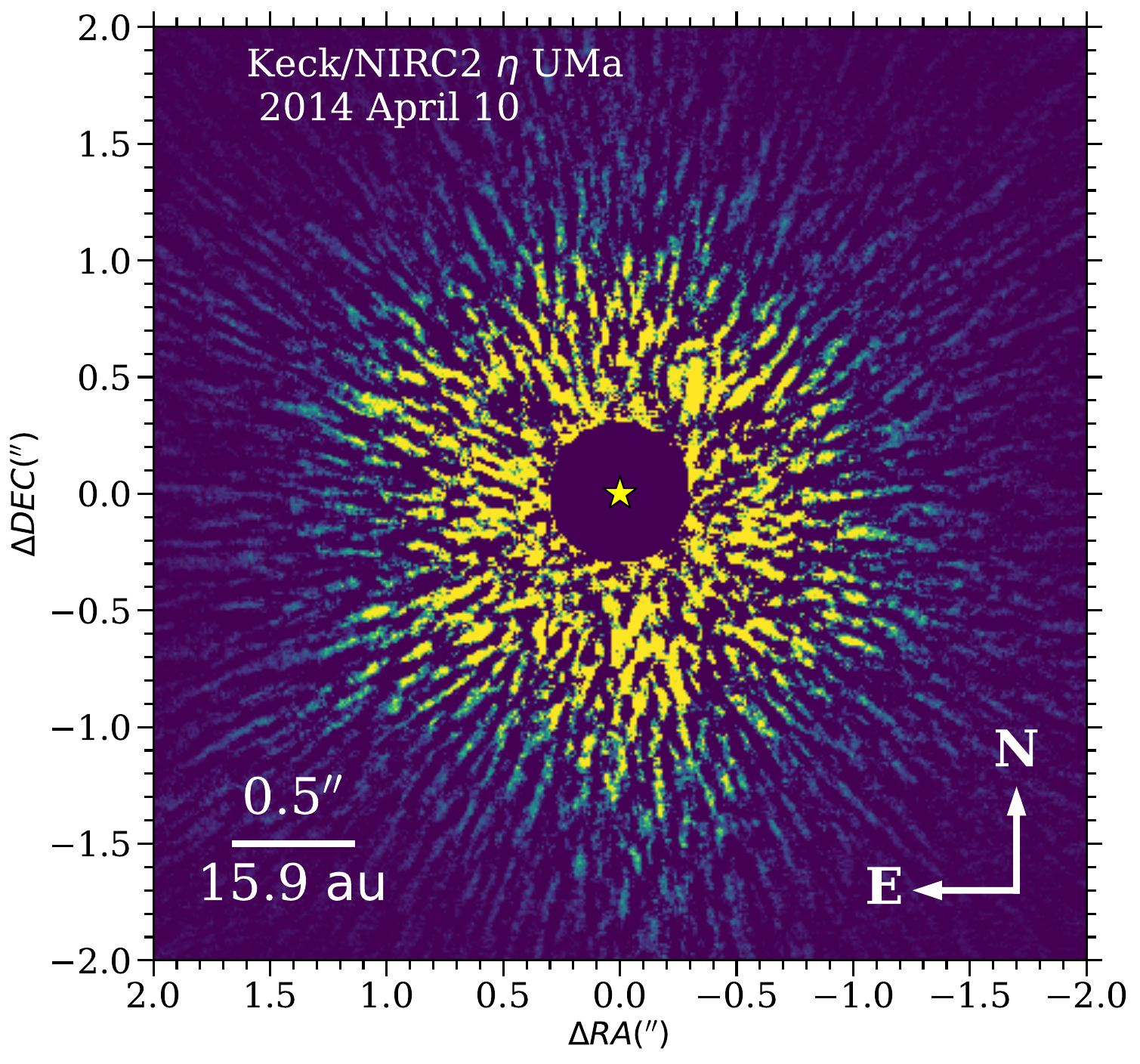}
   \vspace{-0.3in}
  \caption{PSF-subtracted Keck/NIRC2 $K_{\rm s}$ band images for $\alpha$ Cep and $\eta$ UMa.}
  \label{fig:finalimageskeck}
\end{figure}

\subsection{Reduced Images and Contrast Limits}
 Figure \ref{fig:finalimages} shows our final PSF-subtracted, wavelength-collapsed CHARIS images.   For three of the four CHARIS targets, the combination of ADI and SDI yield a nearly flat residual floor over $\rho$ $\sim$ 0\farcs{}3--1.0, with $\sim$ a factor of 10 larger residuals closer to the coronagraph mask edge (i.e. at $\rho$ $\sim$ 0\farcs{}13--0\farcs{}2). 
 Combined with its smaller parallactic angle rotation and poorer raw contrast, HIP 105199's subtraction residuals are brighter.  Visual inspection of each image does not reveal clear evidence for a point source. 

 To assess whether any of the residuals could identify statistically significant signal from a companion or background star, we adopted the standard practice of replacing each pixel with a FWHM-sized aperture sum (i.e. convolving with a top-hat filter sized to the PSF core) and computing the robust standard deviation of convolved pixel values as a function of angular separation \citep[e.g.][]{Marois2008b,Currie2011}.  We adjusted the noise values to account for finite sample sizes following \citet{Mawet2014} such that a 5-$\sigma$ detection would yield a false alarm probability of $\sim$ 2.86 $\times$10$^{-7}$, characteristic of gaussian noise statistics.   We failed to identify any residuals with a greater than 5-$\sigma$ significance and thus conclude that the CHARIS data reveal no companions.

 The final PSF subtracted NIRC2 data (Figure \ref{fig:finalimageskeck}) assess evidence for companions or background stars out to $\approx$ 4\arcsec{}.  Following the same procedures as carried out for CHARIS, we likewise fail to identify any residuals consistent with a $>$ 5-$\sigma$ detection of a companion.  

 Figure \ref{fig:contrast} (top panel) plots each target's 5-$\sigma$ contrast curve.   The SCExAO/CHARIS $\gamma$ Boo data yield our deepest contrasts: $\sim$1.4$\times$10$^{-5}$, $\sim$6$\times$10$^{-6}$,  and 10$^{-6}$ to 4$\times$10$^{-7}$ at 0\farcs{}16, 0\farcs{}25, 0\farcs{}5 to 1\arcsec{}.   The low-wind-effect plagued $\alpha$ Cep dataset resulted in contrasts poorer by about a factor of 5 (10) at 0\farcs{}5 (0\farcs{}15--0\farcs{}25).  Contrasts for the $\eta$ UMa and $\epsilon$ UMa data, which suffered from milder low-wind effect and/or a visible wind-driven halo, were intermediate: $\sim$ 3--5$\times$10$^{-5}$, 1--1.4$\times$10$^{-5}$, and 2$\times$10$^{-6}$ to 5$\times$10$^{-7}$ at 0\farcs{}16, 0\farcs{}25, 0\farcs{}5 to 1\arcsec{}.  The Keck data for $\alpha$ Cep and $\eta$ UMa yielded similar 10$^{-5}$--10$^{-6}$ contrasts but at 1--2 \arcsec{} instead of 0\farcs{}15--1 \arcsec{}.  
 
 The displayed contrast curves assume L dwarf spectra.  Unlike for extreme AO IFS instruments focused on a single band \citep[e.g. GPI;][]{Macintosh2014}, we found little to no difference ($\lesssim$10\%) in contrasts assuming an L dwarf spectrum vs. a T dwarf spectrum.  The reason for this contrast insensitivity is likely that both L and T dwarfs have multiple peaks and troughs across CHARIS's broad spectral bandpass, so that channel-to-channel flux density differences average out.  In contrast, for standard $H$- and $K$-band filters, L dwarfs exhibit a relatively flat, broad signal, while T dwarfs show a narrower peak due to methane absorption. 

\subsection{Companion Mass Limits and V-Band Magnitude Limits}
To convert our contrast limits into companion mass limits, we use the \texttt{species} high-contrast imaging analysis package \citep{Stolker2023}.  We adopt a flat response between 1.1 $\mu m$ and 2.4 $\mu m$ for the CHARIS filter and compute the companion brightness for a given mass and age using spectra primarily from the ATMO-CEQ models for substellar objects \citep{Phillips2020} and the AMES-DUSTY atmosphere models coupled to the Baraffe evolutionary models for more massive objects \citep{Baraffe2003,Allard2012}.  Except for $\eta$ UMa, the targets are sufficiently old that objects in the planetary regime ($\lesssim$20--25 $M_{\rm Jup}$) are generally cool -- with spectral types later than the L/T transition and clouds sequestered well below their photospheres.   As shown later, the ATMO and DUSTY models yield comparable predictions for our V-band contrast limits.

\begin{figure}[ht!]
   \centering
\includegraphics[width=0.5\textwidth,clip]{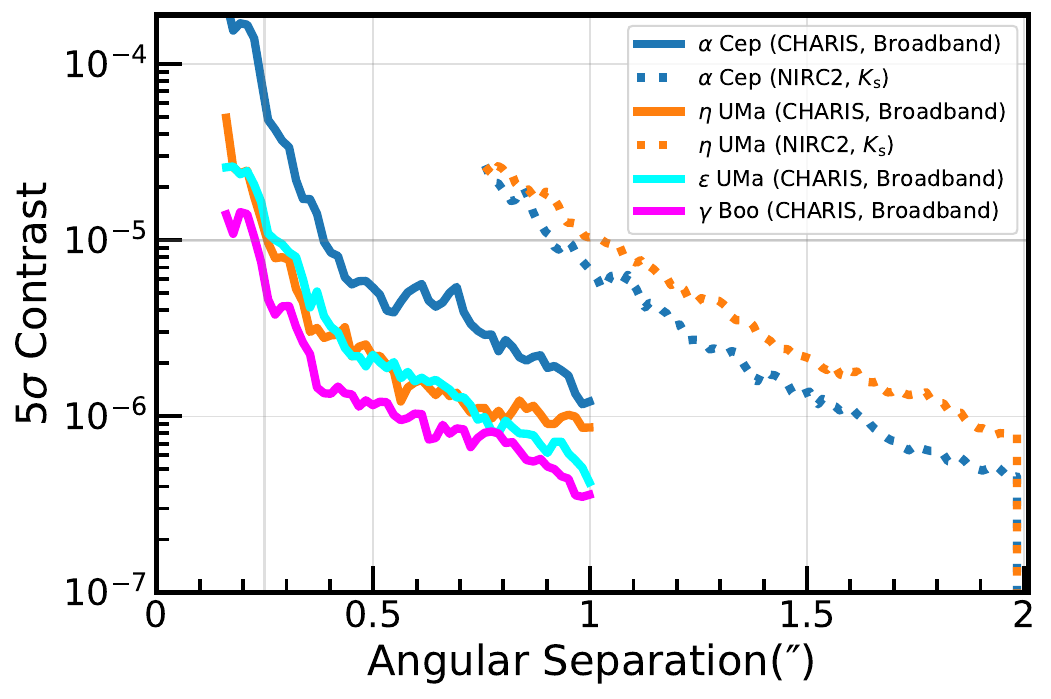}\\
\includegraphics[width=0.5\textwidth,clip]{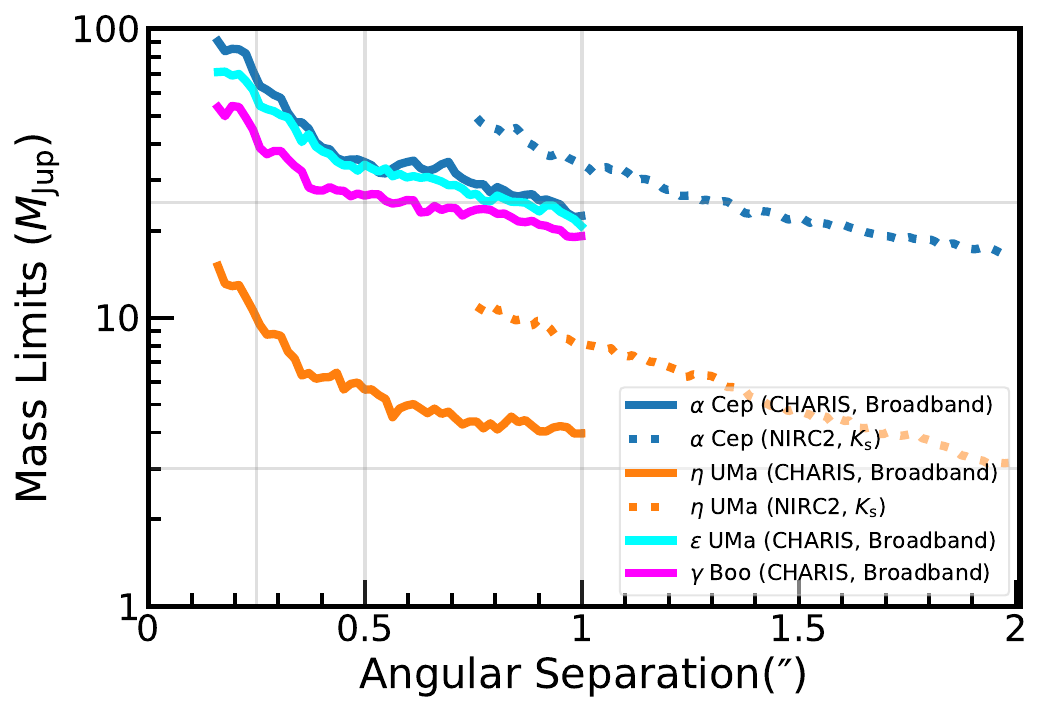}
   \vspace{-0.25in}
 \caption{(Top) The 5-$\sigma$ contrast curves for our targets derived from SCExAO/CHARIS (solid lines) and Keck/NIRC2 (dotted lines).   Light gray vertical lines denote angular separations of 0\farcs{}25, 0\farcs{}5, 1\arcsec{}, 1\farcs{}5, and 2\arcsec{}; horizontal lines denote contrasts of 10$^{-4}$, 10$^{-5}$, and 10$^{-6}$. (Bottom) Mass limits given our 5-$\sigma$ contrast limits and mappings from contrast to mass from \citet{Phillips2020}.  The light gray horizontal lines for the mass limit panel correspond to 3 $M_{\rm Jup}$ and 25 $M_{\rm Jup}$, roughly the mass of $\beta$ Pic d  and the estimated upper limit of planet masses around A stars \citep[see][]{Gibbs2026,Sutlieff2026,Currie2023a,Currie2023b,Currie2026a}.}
  \label{fig:contrast}
\end{figure}

Most sample stars have intermediate ages at which jovian planets are cold and faint in the optical and near-IR \citep[e.g.][]{Skemer2014,Currie2023b}.  Consequently, our data are typically sensitive to brown dwarfs at $\rho$ $\sim$ 0\farcs{}15--0\farcs{}25 and only the most massive planets beyond 0\farcs{}5 (Figure \ref{fig:contrast}, bottom panel). Due to $\eta$ UMa's youth, our limits reach $\sim$ 10 $M_{\rm Jup}$ near CHARIS's inner working angle and $\sim$ 4 $M_{\rm Jup}$ beyond 0\farcs{}5, approaching the dynamical mass of $\beta$ Pic d \citep[$\approx$2--4 $M_{\rm Jup}$][]{Gibbs2026,Sutlieff2026}.  

To estimate the maximum relative brightness of any undetected companion in the Roman Coronagraph's Band 1 due to thermal emission, we computed synthetic photometry in $V$ band for each point in our mass limit curves and subtracted this apparent magnitude from the star's $V$ band photometry to yield a 5-$\sigma$ $V$ band contrast.  Recent models predict that substellar companions have significant sodium absorption strongly overlapping with the Roman Coronagraph's Band 1 centered on 575 nm \citep{LacyBurrows2020,Bovie2025,Currie2026a,Currie2025a,ElMorsy2025}.   $V$ band -- centered on 550 nm -- is a broad bandpass slightly bluer than Band 1 more weakly affected by sodium absorption.  Thus, our contrast limits derived from thermal emission in Band 1 may be conservative.

Planets also reflect starlight, and this reflected-light component may dominate over thermal emission for cool companions typical of older and/or lower-mass gas giants and brown dwarfs \citep[e.g.][]{Currie2023b}.  We coarsely estimate reflected-light $V$-band limits assuming that the unseen companions are jovian planets with a radius of $r_{\rm p}$ and semimajor axis of $a_{\rm p}$ such that
\begin{equation}
\Delta V \approx 1.2\times10^{-9}\times(\frac{r_{\rm p}}{1 R_{\rm Jup}})^{2}(\frac{a_{\rm p}}{5.2 au})^{-2},
\end{equation}
assuming a Jupiter-like albedo \citep{Currie2023b}.  The larger of the contrasts from thermal emission and reflected light then defines the $V$-band contrast limits for each target.

\begin{figure}[ht!]
\includegraphics[width=0.5\textwidth,clip]{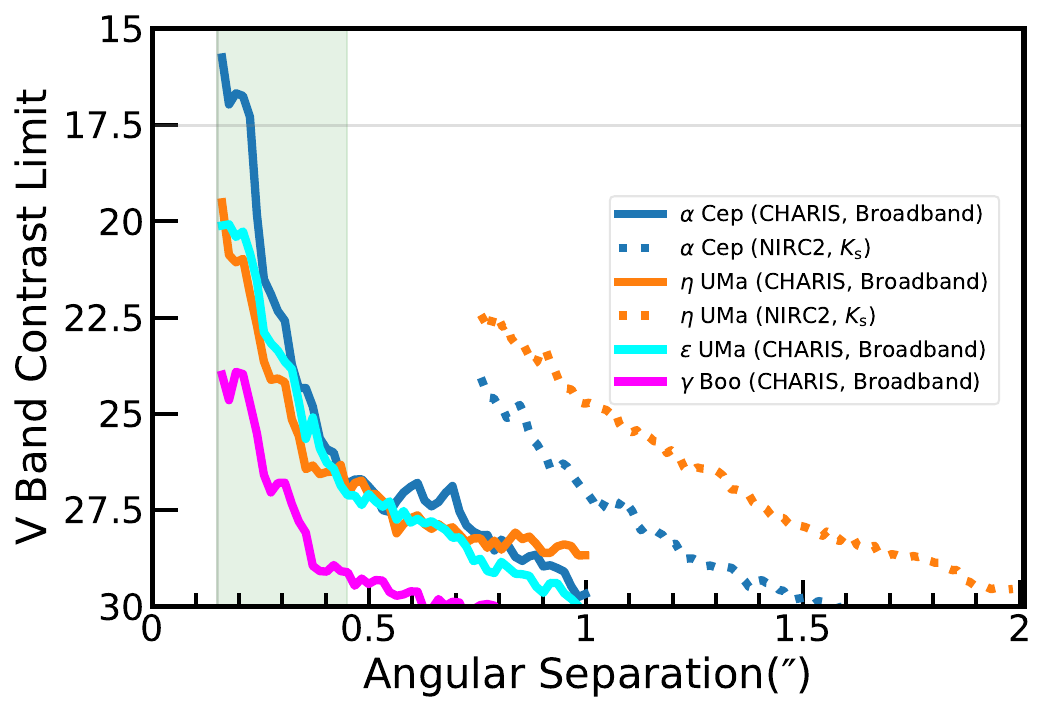}
\includegraphics[width=0.5\textwidth,clip]{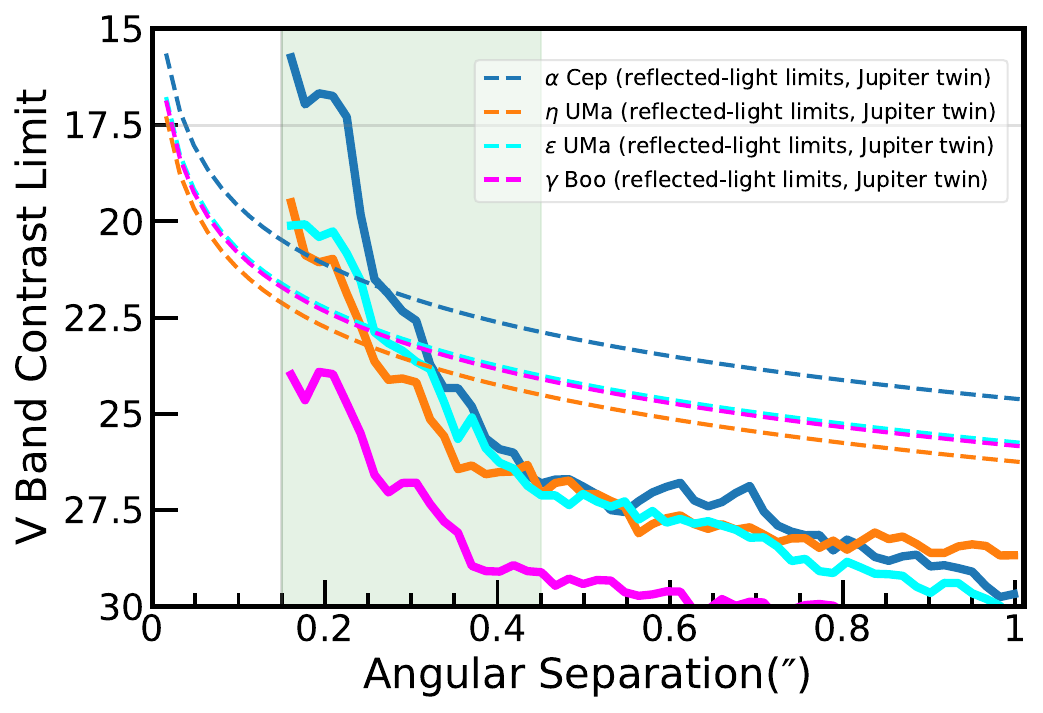}
   \vspace{-0.3in}
  \caption{(Top) Inferred V-band contrast limits, given our 5-$\sigma$ contrast limits and mappings from contrast to $V$-band apparent magnitude from \citet{Phillips2020}.    The shaded region denotes the angular separation range for the Roman Coronagraph's deep-contrast dark hole in Band 1 (575 nm). (Bottom) A focus on V-band contrast limits for the inner arcsecond region with limits for a reflected-light Jupiter twin overplotted as dashed lines.}
  \label{fig:sensitivity}
\end{figure}

Figure \ref{fig:sensitivity} displays our derived V-band contrast limits for thermal emission alone (top panel) and both thermal emission and reflected light (bottom panel).  For three of the four targets -- $\eta$ UMa, $\epsilon$ UMa, and $\gamma$ Boo -- 
our CHARIS data disfavor the existence of companions with $V$ band contrasts brighter than 10$^{-8}$, 10$^{-9}$, and $10^{-10}$ at 0\farcs{}15, 0\farcs{}3, and 1 $\arcsec{}$.  For $\alpha$ Cep, our limits are about a factor of 10 brighter\footnote{Despite these poorer limits, we are still able to rule out a companion precluding TTR5-level contrast over a wide parameter space.  The $V$ band contrast limits for $\alpha$ Cep are orders of magnitude below the near-IR limits reported in \citep{Hom2026}. Only in the innermost
$\rho \sim 0\farcs15$--$0\farcs25$ interval do our limits not yet reach the 10$^{-7}$ threshold: we have no positive evidence for a TTR5-violating companion anywhere on the $\alpha$ Cep field of view.}.   
Except for a narrow $\rho$ $\sim$ 0\farcs{}15--0\farcs{}25 range for $\alpha$ Cep, our current limits directly validate required TTR5 limits (10$^{-7}$) for all targets within the Band 1 dark hole region.
Furthermore, typical contrast limits beyond 0\farcs{}3 are $\lesssim$ 10$^{-9}$, below the flux ratio for a Jupiter-like planet around a Sun-like star in reflected light \citep{Currie2023b}.  Similarly, archival Keck data disfavor the presence of companions with $\Delta$V $\gtrsim$10$^{-8}$ at wider separations.



\section{Scheduling Roman Coronagraph Observations}
\label{sec:schedule}
The non-detection of companions to our sample raises our confidence that they may be suitable PSF reference stars.  
To further investigate our sample's suitability as Roman Coronagraph PSF reference stars, we compare their observability windows and pitch angles to those for HIP 71618 and HIP 54515 during the first year of coronagraph operations.
This analysis updates and extends the study from \citet{Currie2025a,Currie2025b,Currie2026b}, which included the now-discarded $\beta$ Leo and omitted $\alpha$ Cep and $\epsilon$ UMa.  For the relevant observational time frame, we assume an August 30, 2026 launch date -- the current estimate as of 12 August 2026 -- followed by a 3-month commissioning phase before HIP 71618 and HIP 54515 observations.  Thus, we consider observations over one year starting on 1 December 2026.

\begin{figure*}[ht!]
   \includegraphics[width=1\textwidth,clip]
   {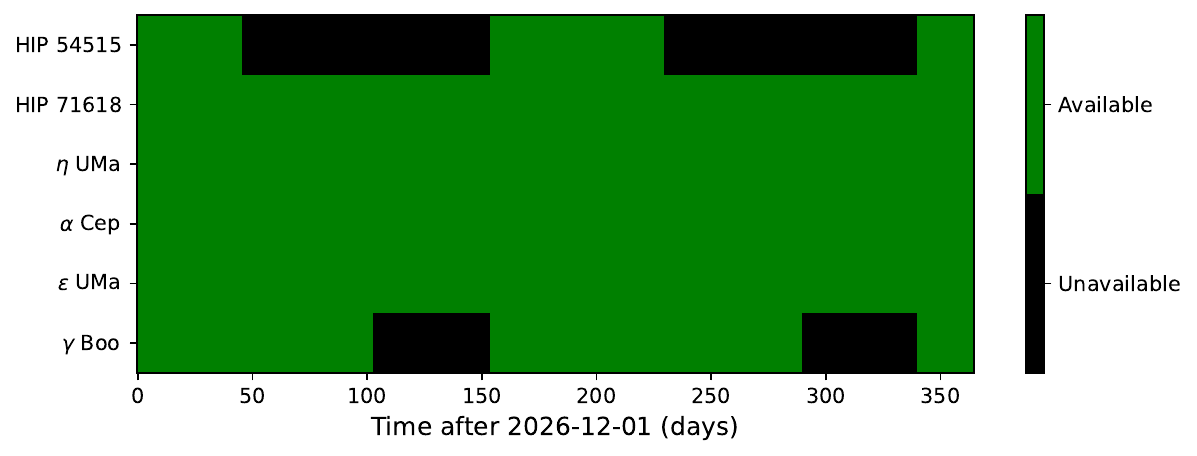}
   \vspace{-0.3in}
  \caption{
Keep-out maps for HIP 54515, HIP 71618, and the candidate PSF reference stars for one calendar year starting with 1 December 2026.
  }
  \label{fig:keepout}
\end{figure*}

Roman Coronagraph observations are restricted to times where the spacecraft's pointing is at an angle of 54--126$^{o}$ with respect to the Sun. Figure \ref{fig:keepout} shows ``keepout maps" identifying where such observations are impossible for our targets and PSF reference stars.  Being within the CVZ, the target HIP 71618 and reference stars $\eta$ UMa, $\alpha$ Cep, and $\epsilon$ UMa have minimal timing restrictions, while $\gamma$ Boo is only unobservable in two $\sim$ 25-day windows.   Scheduling is slightly more restrictive for HIP 54515 given its further distance from the CVZ, resulting in the target being visible $\sim$50\% of the time.

\begin{figure}[ht!]
   \includegraphics[width=1\columnwidth,clip]
   {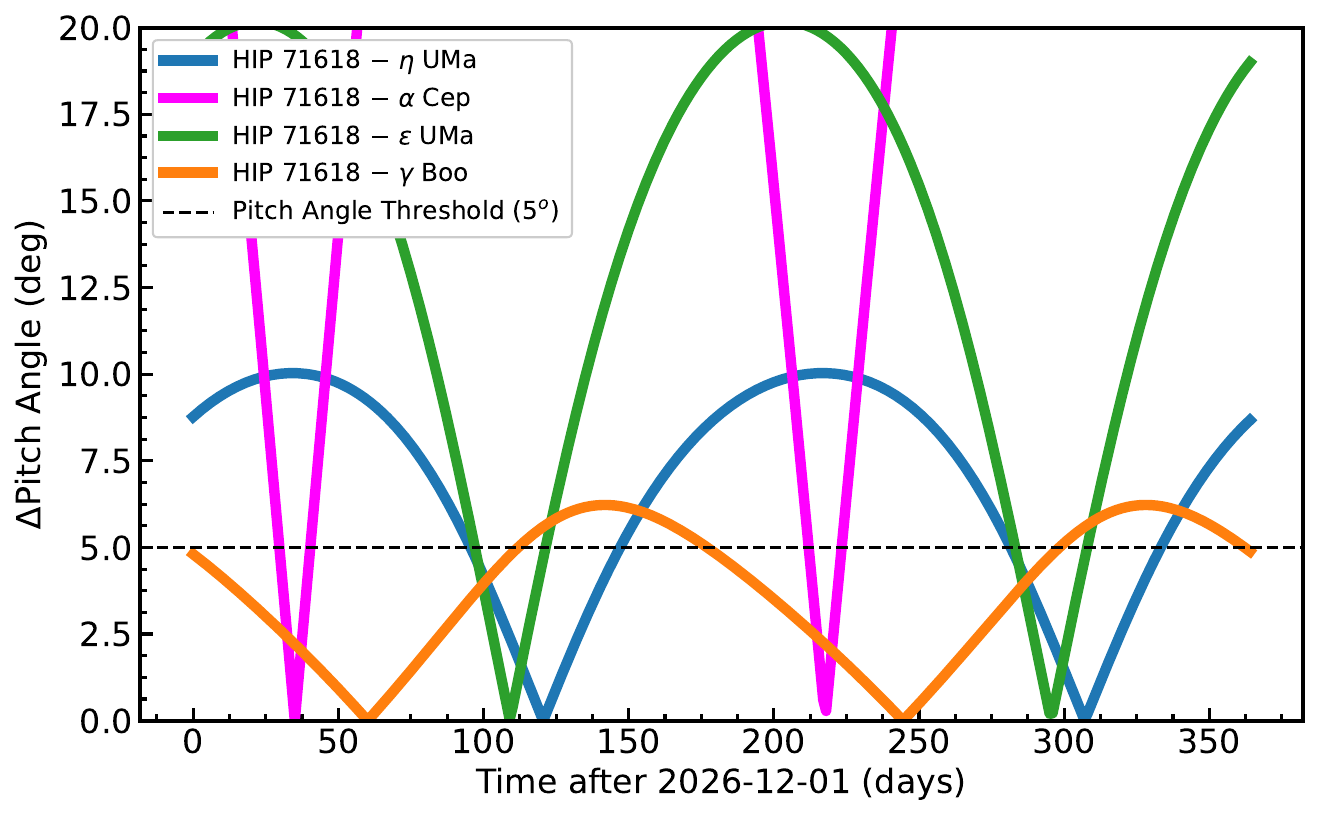}\\
      \includegraphics[width=1\columnwidth,clip]
   {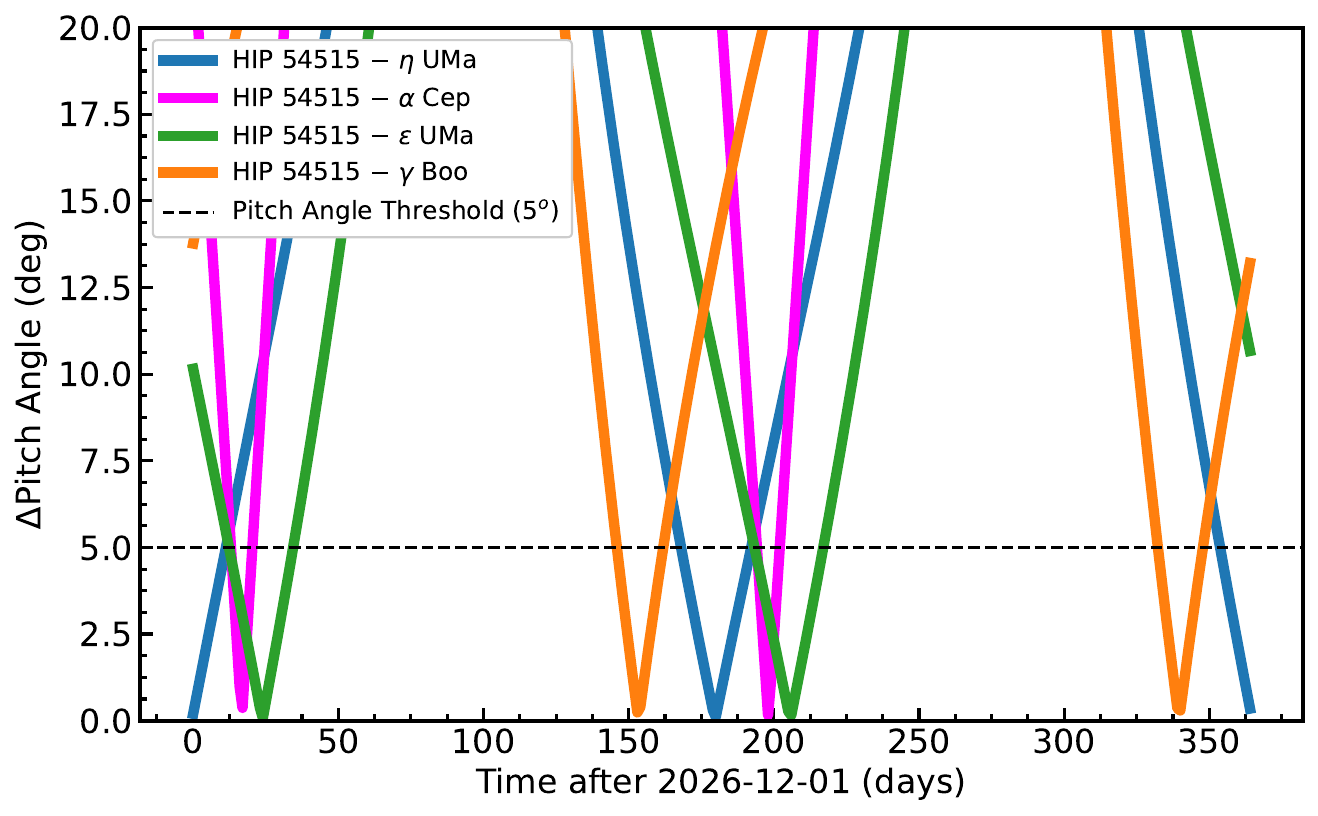}
   \vspace{-0.1in}
  \caption{
Differences in the pitch angle between targets (HIP 71618, HIP 54515) and candidate PSF reference stars analyzed in this paper.  The horizontal dashed line denotes the maximum angle difference nominally acceptable for technology demonstration observations.
  }
  \label{fig:pitchangle}
\end{figure}

Figure \ref{fig:pitchangle} tracks the difference in pitch angles between HIP 71618 and HIP 54515 with the four candidate reference stars.  At least one of the candidate PSF references has $\Delta$ Pitch Angle $\le$ 5$^{o}$ with HIP 71618 for the entirety of the year. For HIP 54515, stars in the CVZ we study generally have much larger pitch angle differences.  However, pitch angle differences are less than $\sim$5 $^{o}$ for at least one reference star over two 75-day blocks.

Table \ref{cgi_schedule} summarizes the scheduling opportunities for HIP 71618, HIP 54515, and compatible reference stars.  In particular, HIP 71618 and multiple reference stars are easily targetable in the December 2026 - February 2027 window likely after commissioning but prior to HLTDS priority time.  HIP 54515 has a small targetable window in this range and a larger one immediately after (March-May 2027) when paired with reference stars  $\eta$ UMa, $\alpha$ Cep, and $\epsilon$ UMa.

\begin{deluxetable*}{lllllllll}[ht]
     \tablewidth{0pt}
    \tablecaption{Earliest Roman Coronagraph Scheduling Windows for Target+Reference Star Pairs}
    \tablehead{\colhead{} & \colhead{$\eta$ UMa} & \colhead{$\alpha$ Cep} & \colhead{$\epsilon$ UMa} & \colhead{$\gamma$ Boo}}
    \startdata
    HIP 71618 & mid Mar. 2027- late Apr. 2027$^{a}$ & early-mid Jan. 2027 & Mar. 2027 & Dec 1 2026-mid Mar. 2027 \\
    HIP 54515 & early-mid Dec. 2026 & mid-late Dec. 2026 &late Dec. 2026 - early Jan 2027& early-mid May 2027
    \enddata
    \tablecomments{Visibility is determined by a combination of the keepout maps and $\Delta$pitch angles. a) If observations can be taken between launch and the assumed 3 month commissioning phase, $\eta$ UMa has an additional window until about November 10 2026. }
    \label{cgi_schedule}
\end{deluxetable*}

\section{Discussion}
Using deep SCExAO/CHARIS high-contrast integral field spectroscopy and archival Keck/NIRC2 high-contrast imaging, we  analyze four candidate Roman Coronagraph PSF reference stars in the Continuous Viewing Zone.   Our data fail to identify any 5 $\sigma$-significant signals consistent with a companion that could degrade the ability of CGI to dig a deep-contrast dark hole.  Our contrast limits strongly disfavor the presence of brown dwarfs at $\rho$ $\sim$ 0\farcs{}15--0\farcs{}25 and massive planets at wider separations.  Translated into the Roman Coronagraph's Band 1 (575 nm), these limits  disfavor the existence of a companion with a $V$-band contrast $\gtrsim$ 10$^{-8}$, 10$^{-9}$, and $10^{-10}$ at 0\farcs{}15, 0\farcs{}3, and 1$\arcsec{}$ around three of the four targets.  Limits for $\alpha$ Cep are a factor of 10 poorer.  

  For three of our targets, our contrast limits eliminate the phase space of contaminants capable of corrupting the Coronagraph's dark hole down to the TTR5 baseline of 10$^{-7}$.  Over most of that phase space ($\rho$ $\gtrsim$ 0\farcs{}4), we reach limits comparable to the reflected-light contrast of a Jupiter twin.  While $\alpha$ Cep's limits retain a small phase space where a $\gtrsim$10$^{-7}$-contrast companion may yet exist, higher-quality second-epoch CHARIS observations may resolve this issue.  Prior interferometry likewise finds no bright binary around three of the targets \citep{Hom2026}, and the reported close binary around the fourth ($\gamma$~Boo) is plausibly spurious.   Unless these targets have faint substellar companions within $\rho$ $\sim$ 0\farcs{}15--0\farcs{}25, they likely lack background stars or companions whose presence could corrupt the Roman Coronagraph’s dark hole digging sufficiently to preclude the detection of reflected-light planets.  Broadly speaking, available data thus support all four as suitable PSF reference candidates for the technology demonstration.  

 Our study's keep-out and pitch-angle analyses further establish that these targets are practically schedulable as references for key tech demo targets HIP 71618 and HIP 54515. All three CVZ stars -- $\eta$~UMa, $\alpha$~Cep, and $\epsilon$~UMa -- can be co-observed with HIP~71618 with minimal timing restrictions: at least one maintains $\Delta$(pitch)~$\le 5^{\circ}$ with HIP~71618 throughout the first year of coronagraph operations. $\gamma$~Boo supplements this coverage outside two $\sim$25-day windows.  HIP~54515 lies farther from the CVZ and thus is visible $\sim$50\% of the time.  It pairs with compatible references in a short window near launch and a broader interval in March--May 2027. Assuming the current August~30, 2026 launch and a three-month commissioning phase, the earliest opportunity to achieve TTR5 on HIP 71618 is the December~2026--February~2027 window, when this target and multiple references are simultaneously accessible after commissioning but before High-Latitude Time-Domain Survey priority time.

These results establish the first pool of extreme-AO-vetted PSF references for the Roman Coronagraph technology demonstration, to a depth equal to that required for TTR5 over almost all phase space, orders of magnitude deeper than previously published work, and orders of magnitude beyond the requirement for most stars. 
Delivering this vetting before launch allows $\eta$~UMa, $\epsilon$~UMa, $\gamma$~Boo, and -- outside a narrow $\rho \sim 0\farcs15$--$0\farcs25$ range -- $\alpha$~Cep to be slotted against HIP~71618 and HIP~54515 from the opening of the technology demonstration phase. This broad approach, using ground-based extreme-AO integral field spectroscopy to retire reference-star risk for reflected-light coronagraphy, is transferable to reference star selection for the Habitable Worlds Observatory.

\begin{acknowledgments}

\indent The authors acknowledge the very significant cultural role and reverence that the summit of Maunakea holds within the Hawaiian community.  We are most fortunate to have the opportunity to conduct observations from this mountain.

\indent This research has made use of the Keck Observatory Archive (KOA), which is operated by the W. M. Keck Observatory and the NASA Exoplanet Science Institute (NExScI), under contract with the National Aeronautics and Space Administration.\\
\indent The development of SCExAO was supported by JSPS (Grant-in-Aid for Research \#23340051, \#26220704 \& \#23103002), Astrobiology Center of NINS, Japan, the Mt Cuba Foundation, and the director's contingency fund at Subaru Telescope.  CHARIS was developed under the support by the Grant-in-Aid for Scientific Research on Innovative Areas \#2302.  SCExAO’s adaptive optics loops and high-speed data acquisition are handled by  the CACAO package, which is supported by NSF award 2410616. \\
\indent This work is generously supported by National Science Foundation (NSF) Astronomy and Astrophysics grant \#2408647 and NASA-Keck Strategic Mission Support Proposal.  We are grateful for the continued valuable work and expert guidance of NSF and NASA-Keck personnel, especially in challenging times. \\
\end{acknowledgments}


\bibliographystyle{aasjournal}
\bibliography{bibliography}

\end{document}